\documentclass[letterpaper,10pt,conference]{ieeeconf}
\IEEEoverridecommandlockouts
\usepackage[T1]{fontenc}
\usepackage{amsmath,amssymb}
\usepackage{balance}
\usepackage{tikz}
\usetikzlibrary{arrows.meta}
\usepackage[hidelinks]{hyperref}
\hypersetup{pdftitle={Generic Local Identifiability Does Not Imply Generic Global Identifiability in Linear Networks},
 pdfauthor={Anders Hansson},
 pdfsubject={Draft for IEEE Control Systems Letters},
 pdfkeywords={identification, linear networks, local identifiability, global identifiability}}
\newtheorem{theorem}{Theorem}
\newtheorem{proposition}{Proposition}
\newtheorem{lemma}{Lemma}
\newcommand{\diag}{\operatorname{diag}}
\newcommand{\aicite}{\footnote{Drafting assistance: OpenAI Codex \cite{codex}.}\space}

\title{\LARGE\bf
Generic Local Identifiability Does Not Imply
Generic Global Identifiability in Linear Networks}
\author{Anders Hansson, \textit{Senior Member, IEEE}%
\thanks{Anders Hansson is with the Division of Automatic Control,
Link\"oping University. Email:
\href{mailto:anders.g.hansson@liu.se}{\texttt{anders.g.hansson@liu.se}}.}}

\begin{document}
\maketitle
\thispagestyle{empty}
\pagestyle{empty}

\begin{abstract}
We give a counterexample to the implication from generic local to
generic global identifiability for linear networks with a fixed
topology, independently unknown edge coefficients, and selected
excited and measured nodes. A six-node network with nine unknown
coefficients has exactly two compatible parameter vectors generically.
Adding six excited and measured nodes preserves this ambiguity while
making the added coefficients directly recoverable. The resulting
twelve-node structure has forty-two independent unknown coefficients
and zero diagonal. For every positive bound on the largest eigenvalue
magnitude, a nonempty open set of parameters has two distinct compatible
networks satisfying that bound. This holds for both real and complex
coefficients. Other open regions have exactly one admissible solution.
The proof uses a rational transformation exchanging the two solutions
and a common limiting matrix whose fourth power vanishes. Exact
rational-number certificates verify an explicit pair. The construction
also extends algebraically when the unknown coordinates are whole
rational transfer functions. Genericity in prescribed numerator and
denominator coefficient families, and the requirement that every
transfer function vanish at infinite frequency, remain separate questions.
\end{abstract}

\begin{keywords}
Identification, network analysis and control, linear systems.
\end{keywords}

\section{Introduction}\label{sec:intro}

A linear network relates node signals through coefficients on prescribed
directed edges. Inputs excite selected nodes, and measurements record
selected node signals. Network identification seeks to recover the
unknown edge coefficients from the resulting input-output relation.\aicite

Local identifiability means that the coefficients are unique among
sufficiently close admissible choices. Global identifiability requires
uniqueness among all admissible choices. For scalar coefficients,
a property is generic if it holds outside a set of zero volume.
A derivative test for local uniqueness can leave a second, isolated
explanation of the same input-output relation.

Legat and Hendrickx developed a generic local identifiability test for
partial excitation and measurement and asked whether local and global
identifiability can differ in this setting \cite{LH20}. The two notions
coincide when all nodes are excited or all nodes are measured
\cite{LH24}. A further equivalence result holds for a specified class
of two-part networks with known internal coefficients \cite{LH23}.
The general question remains open in the
discussion of \cite[Sec.~II-C]{LH24} and
\cite[Sec.~12.2]{L25}. The latter reports an unsuccessful search for a
counterexample with up to five unknown edges, subject to specified
structural restrictions.

This letter answers the question for the cited scalar formulation.
The topology and node selections are fixed, every permitted edge
has an independent unknown coefficient, and excitation and measurement
are both partial. A six-node construction has exactly two compatible
vectors generically. Six added excited and measured nodes preserve the
ambiguity and make both spectral radii, the largest eigenvalue
magnitudes, arbitrarily small. The resulting structure has forty-two
independent coefficients. For every positive spectral-radius bound,
ambiguity occurs on open subsets of both the full complex and real
parameter spaces. Both alternatives satisfy the same bound, including
the restriction in \cite{LH24}. The node and edge counts are
constructive upper bounds; minimality remains open.

The scalar setting allows direct feedthrough, meaning an instantaneous
dependence between connected node signals. Constant real coefficients
therefore also give transfer-function examples. We further use the
algebraic framework of \cite{HMA26}, where each coordinate is a whole
rational transfer function and genericity excludes polynomial equations
in these functions. This differs from varying the numerical numerator
and denominator coefficients at a prescribed dynamic order.
Such coefficient families, and strictly proper functions that vanish
at infinite frequency, require separate analysis.

Section~\ref{sec:setting} states the admissible classes and the precise
identifiability definitions. Sections~\ref{sec:base} and
\ref{sec:augmentation} develop the two constructions.
Section~\ref{sec:main} proves the spectral-radius result and gives an
explicit pair with rational entries. Section~\ref{sec:conclusion} extends the result
to rational transfer functions under an algebraic definition of generic
identifiability.

\section{Network Description and Identifiability}
\label{sec:setting}

A transfer function describes a linear relation between signals in the
frequency domain. We follow the scalar formulation of
\cite[Secs.~I and III]{LH20} and \cite[Sec.~III-B]{LH24}: each edge
coefficient represents a transfer-function value at a fixed frequency and
can therefore be complex.\aicite

Let \(n\) be the number of nodes, numbered \(1,\ldots,n\), and let \(E\)
be a prescribed set of ordered pairs \((i,j)\), where
\(i,j\in\{1,\ldots,n\}\) and \(i\ne j\). An edge
\((i,j)\) carries the signal from node \(j\) to node \(i\).
Write \(m=|E|\), fix an ordering of these edges, and collect their
coefficients in a vector \(\theta\in\mathbb C^m\). The network matrix
\(G(\theta)\in\mathbb C^{n\times n}\) places these coefficients in the
positions in \(E\) and has zeros elsewhere. Each edge has its own
independent parameter. In particular, \(G(\theta)\) has zero diagonal.

Let \(n_B\) and \(n_C\) be the numbers of excited and measured nodes.
The fixed matrix \(B\in\{0,1\}^{n\times n_B}\) has distinct columns of
the identity matrix, and \(C\in\{0,1\}^{n_C\times n}\) has distinct rows
of the identity matrix. Thus \(B\) inserts inputs at the excited nodes
and \(C\) selects the measured signals. At the chosen frequency, let
\(w\in\mathbb C^n\), \(r\in\mathbb C^{n_B}\), and
\(y\in\mathbb C^{n_C}\) be the complex amplitudes of the node signals,
inputs, and measurements, respectively. The network equations are
\begin{equation}\label{eq:network}
 w=G(\theta)w+Br,\qquad y=Cw.
\end{equation}
For a positive integer \(d\), let \(I_d\) denote the \(d\times d\)
identity matrix. We call \eqref{eq:network} well posed if
\(I_n-G(\theta)\) is invertible. In that case the transfer matrix
from the external inputs \(r\) to the measurements \(y\), evaluated
at the chosen frequency, is
\begin{equation}\label{eq:observation}
 \Phi(\theta)=C(I_n-G(\theta))^{-1}B.
\end{equation}
Identification concerns recovery of \(\theta\) from \(\Phi(\theta)\).
The fiber at a transfer matrix is the set of admissible parameter
vectors producing it.

Define the set of well-posed parameters by
\[
 \Omega=\{\theta\in\mathbb C^m:\det(I_n-G(\theta))\ne0\}.
\]
For a matrix, let \(\rho\) denote its spectral radius. For
\(0<\varepsilon\le1\), define the more restricted set
\begin{equation}\label{eq:domain}
 \Omega_\varepsilon
   =\{\theta\in\mathbb C^m:\rho(G(\theta))<\varepsilon\}.
\end{equation}
The strict inequality implies \(\Omega_\varepsilon\subseteq\Omega\).
Openness and volume are understood by identifying \(\mathbb C^m\)
with \(\mathbb R^{2m}\) through real and imaginary parts.
Both parameter sets are open in this topology.

Let \(\mathcal D\) be either \(\Omega\) or \(\Omega_\varepsilon\).
The network is \emph{locally identifiable at \(\theta\in\mathcal D\)}
if some neighborhood \(N\) of \(\theta\) satisfies the following
implication for every alternative vector \(\widetilde\theta\):
\[
 \widetilde\theta\in N\cap\mathcal D,\quad
 \Phi(\widetilde\theta)=\Phi(\theta)
 \quad\Longrightarrow\quad \widetilde\theta=\theta.
\]
It is \emph{globally identifiable at \(\theta\) in \(\mathcal D\)}
if the same implication holds for every
\(\widetilde\theta\in\mathcal D\). Thus any competing network must satisfy
the same admissibility restriction.

The structure is generically locally or globally identifiable in
\(\mathcal D\) if the corresponding property holds outside a subset of
\(\mathcal D\) of Lebesgue measure zero, that is, zero volume.
We use the complex Jacobian, the matrix of derivatives of the entries
of \(\Phi\) with respect to the complex coordinates of \(\theta\).
All ranks are over \(\mathbb C\).
We prove generic local identifiability by showing that this Jacobian
has column rank \(m\) except where a nonzero polynomial vanishes.
Such an exceptional set has zero volume in \(\mathbb R^{2m}\).
For the rational map \(\Phi\), complex rank \(m\) gives real rank \(2m\)
after separating real and imaginary parts. Applying the inverse function
theorem to \(m\) suitable complex entries of \(\Phi\) then gives local
uniqueness. Conversely, a nonempty open
set of parameters with distinct admissible alternatives suffices to
disprove generic global identifiability.
This conclusion alone does not say that almost every admissible
parameter vector has an alternative.

The generic claims in Sections~\ref{sec:base}--\ref{sec:main} refer to
the independent scalar parameters. Section~\ref{sec:conclusion}
considers independent rational transfer functions as the unknowns.

\section{A Six-Node Ambiguity}
\label{sec:base}

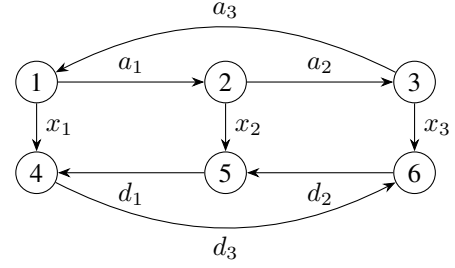
\begin{figure}[t]
\centering
\begin{tikzpicture}[>=Stealth,
 vertex/.style={circle,draw,minimum size=5.5mm,inner sep=1pt}]
\node[vertex] (n1) at (0,1.2) {1};
\node[vertex] (n2) at (2.5,1.2) {2};
\node[vertex] (n3) at (5,1.2) {3};
\node[vertex] (n4) at (0,0) {4};
\node[vertex] (n5) at (2.5,0) {5};
\node[vertex] (n6) at (5,0) {6};
\draw[->] (n1) -- node[above] {\(a_1\)} (n2);
\draw[->] (n2) -- node[above] {\(a_2\)} (n3);
\draw[->] (n3) to[bend right=27] node[above] {\(a_3\)} (n1);
\draw[->] (n5) -- node[below] {\(d_1\)} (n4);
\draw[->] (n6) -- node[below] {\(d_2\)} (n5);
\draw[->] (n4) to[bend right=27] node[below] {\(d_3\)} (n6);
\draw[->] (n1) -- node[right] {\(x_1\)} (n4);
\draw[->] (n2) -- node[right] {\(x_2\)} (n5);
\draw[->] (n3) -- node[right] {\(x_3\)} (n6);
\end{tikzpicture}
\caption{The six-node construction. Nodes \(1,2,3\) are excited;
nodes \(4,5,6\) are measured. Each labeled edge has an independent
unknown coefficient.}
\label{fig:base}
\end{figure}

Consider nine independent complex parameters \(a_i,d_i,x_i\), indexed by
\(i\in\{1,2,3\}\). In this section and in formulas using these parameters,
subscripts are cyclic: \(i+1=1\) when \(i=3\), and \(i-1=3\) when \(i=1\).\aicite
Let \(\diag\) form a diagonal matrix from its arguments, and define
\begin{equation}\label{eq:blocks}
\begin{split}
 A&=\begin{pmatrix}0&0&a_3\\a_1&0&0\\0&a_2&0\end{pmatrix},\qquad
 D=\begin{pmatrix}0&d_1&0\\0&0&d_2\\d_3&0&0\end{pmatrix},\\
 X&=\diag(x_1,x_2,x_3),
\end{split}
\end{equation}
In block matrices, zero blocks have the dimensions determined by their
positions. Set
\begin{equation}\label{eq:base}
 G_b=\begin{pmatrix}A&0\\X&D\end{pmatrix},\quad
 B_b=\begin{pmatrix}I_3\\0\end{pmatrix},\quad
 C_b=\begin{pmatrix}0&I_3\end{pmatrix}.
\end{equation}
This is \eqref{eq:network} with six nodes, the first three excited and
the last three measured. Figure~\ref{fig:base} shows the edges.
The within-part matrices \(A,D\) are unknown. In the separable
networks of \cite{LH23}, these matrices are known and only the
connections between parts are unknown; that equivalence theorem
therefore does not apply here.
Define the products \(\alpha=a_1a_2a_3\) and
\(\delta=d_1d_2d_3\). Then
\[
 \det(I_6-G_b)=(1-\alpha)(1-\delta).
\]
For well-posed parameters, denote the transfer matrix by
\begin{equation}\label{eq:F}
 F=C_b(I_6-G_b)^{-1}B_b=(I_3-D)^{-1}X(I_3-A)^{-1}.
\end{equation}

For nonzero \(x_i\), define the auxiliary quantities
\begin{equation}\label{eq:auxiliary}
 k_i=a_id_i\frac{x_{i+1}}{x_i},\qquad h_i=1+k_i+k_ik_{i+1}.
\end{equation}
Let \(\mathcal D_b\subset\mathbb C^9\) be the domain specified by
\begin{equation}\label{eq:base-domain}
\begin{split}
 x_1x_2x_3&\ne0,\\
 \alpha\delta(1-\alpha)(1-\delta)(1-\alpha\delta)h_1h_2h_3&\ne0.
\end{split}
\end{equation}
Define \(T_b\) on the nine parameters by
\begin{equation}\label{eq:alternative}
 \eta_i=\frac{k_ih_{i+1}}{h_i},\qquad
 (a_i',d_i',x_i')=\frac1{\eta_i}(a_i,d_i,x_i).
\end{equation}
Primes denote the resulting alternative values.

\begin{proposition}\label{prop:base}
The structure \eqref{eq:base} is generically locally identifiable in
\(\Omega\). For every parameter vector in \(\mathcal D_b\), exactly
two well-posed vectors give its transfer matrix \(F\), and both are
locally identifiable. The map \(T_b\) preserves \(\mathcal D_b\) and
is a rational involution: it has rational coordinates and applying it
twice returns the original vector.
\end{proposition}

\begin{proof}
Assume first that \(x_1x_2x_3\ne0\), so that \(F\) is invertible.
Introduce coordinates \(u_i=a_i/x_i\), \(v_i=d_i/x_i\), and \(z_i=1/x_i\).
This change is invertible, with
\(a_i=u_i/z_i\), \(d_i=v_i/z_i\), and \(x_i=1/z_i\).
The inverse transfer matrix is
\begin{equation}\label{eq:Finverse}
\begin{split}
 F^{-1}&=(I_3-A)X^{-1}(I_3-D)\\
 &=\begin{pmatrix}
 z_1+u_3v_3/z_3&-v_1&-u_3\\
 -u_1&z_2+u_1v_1/z_1&-v_2\\
 -v_3&-u_2&z_3+u_2v_2/z_2
 \end{pmatrix}.
\end{split}
\end{equation}
Thus the off-diagonal entries determine \(u_i,v_i\).
Let \(e_i=(F^{-1})_{ii}\) be the diagonal entries. They satisfy
\begin{equation}\label{eq:diagonal}
 e_i=z_i+\frac{u_{i-1}v_{i-1}}{z_{i-1}}.
\end{equation}
With \(u_i,v_i\) fixed, the Jacobian of these three equations with
respect to \((z_1,z_2,z_3)\), denoted by \(J_z\), has determinant
\begin{equation}\label{eq:rankbase}
 \det J_z
 =1-\frac{u_1v_1u_2v_2u_3v_3}{z_1^2z_2^2z_3^2}
 =1-\alpha\delta.
\end{equation}
The six directly determined coordinates and \eqref{eq:rankbase} give
Jacobian rank nine when \(\alpha\delta\ne1\). Inverting an invertible
transfer matrix preserves rank. This proves generic local
identifiability.

Expanding \(h_i\) and using \(k_1k_2k_3=\alpha\delta\) gives
\begin{equation}\label{eq:identities}
\begin{split}
 \eta_i-1&=\frac{\alpha\delta-1}{h_i},\\
 \eta_i+\frac{k_{i-1}}{\eta_{i-1}}&=1+k_{i-1},
 \qquad \eta_1\eta_2\eta_3=\alpha\delta.
\end{split}
\end{equation}
The transformation fixes \(u_i,v_i\) and gives \(z_i'=\eta_i z_i\).
Since \(u_{i-1}v_{i-1}/z_{i-1}=k_{i-1}z_i\), the second identity
in \eqref{eq:identities} preserves \eqref{eq:diagonal}, and hence \(F\).
The first identity makes the alternatives distinct when
\(\alpha\delta\ne1\). Denote the products of the primed cycle
coefficients by \(\alpha'=\prod_{i=1}^3a_i'\) and
\(\delta'=\prod_{i=1}^3d_i'\). The last identity gives
\begin{equation}\label{eq:products}
 \alpha'=\frac1\delta,\qquad \delta'=\frac1\alpha.
\end{equation}
Thus the alternative is well posed on \(\mathcal D_b\), and
\(1-\alpha'\delta'=1-1/(\alpha\delta)\ne0\) gives rank nine there.
Applying \eqref{eq:auxiliary} and \eqref{eq:alternative} to the primed
parameters also gives
\begin{equation}\label{eq:involution}
 h_i'=\frac{h_{i+1}}{k_{i+1}k_{i+2}},\qquad
 \eta_i'=\eta_i^{-1}.
\end{equation}
All factors excluded in \eqref{eq:base-domain} remain nonzero.
Hence \(T_b(\mathcal D_b)=\mathcal D_b\) and \(T_b^2\) is the identity.

It remains to bound the number of solutions. Define \(b_i=u_iv_i\).
Equation~\eqref{eq:diagonal} gives successively
\[
 z_2=e_2-b_1/z_1,\quad
 z_3=e_3-b_2/z_2,\quad z_1=e_1-b_3/z_3.
\]
Substitution and multiplication by the denominators yield
\begin{equation}\label{eq:quadratic}
\begin{split}
0={}&(b_2-e_2e_3)z_1^2\\
 &+(b_1e_3-b_2e_1-b_3e_2+e_1e_2e_3)z_1\\
 &+b_1b_3-b_1e_1e_3.
\end{split}
\end{equation}
Its leading coefficient is \(b_2-e_2e_3=-z_2z_3h_1\), which is
nonzero on \(\mathcal D_b\). There are therefore at most two values
of \(z_1\). Each determines \(z_2,z_3\) and all original
parameters. A candidate with some \(x_i=0\) cannot yield an
invertible \(F\), by \eqref{eq:F}.
Formula~\eqref{eq:alternative} supplies the distinct second solution.
After clearing denominators, the exclusions in \eqref{eq:base-domain}
are zeros of nonzero polynomials: \(a_i=d_i=-1/2,\ x_i=1\) satisfy
all the inequalities. Thus \(\mathcal D_b\) has full volume in
\(\Omega\), completing the generic claim.
\end{proof}

The six-node construction alone does not give two alternatives in
\(\Omega_1\). Indeed, its block triangular form gives
\[
 \rho(G_b)=\max\{|\alpha|^{1/3},|\delta|^{1/3}\}.
\]
If both products have magnitude below one, \eqref{eq:products} makes
the alternative violate the same restriction. The next construction
changes the full network eigenvalues while retaining the ambiguity.

\section{Adding Excited and Measured Nodes}
\label{sec:augmentation}

\begin{figure}[t]
\centering
\vspace*{9pt}
\begin{tikzpicture}[>=Stealth,
 group/.style={draw,align=center,minimum width=2.15cm,
 minimum height=9mm,inner sep=3pt}]
\node[group] (base) at (0,0) {Original nodes \(w_b\)\\Internal matrix \(G_b\)};
\node[group] (added) at (4,0) {Added nodes \(w_c\)\\Internal matrix \(J\)};
\draw[->] (0,1.15) node[above] {\(r_b\)} --
 node[left] {\(B_b\)} (base.north);
\draw[->] (4,1.15) node[above] {\(s\)} -- (added.north);
\draw[->] ([yshift=2mm]base.east) --
 node[above] {\(L_cC_b\)} ([yshift=2mm]added.west);
\draw[->] ([yshift=-2mm]added.west) --
 node[below] {\(B_bK\)} ([yshift=-2mm]base.east);
\draw[->] (base.south) -- (0,-1.05) node[below] {\(y_b=C_bw_b\)};
\draw[->] (added.south) -- (4,-1.05) node[below] {\(w_c\)};
\end{tikzpicture}
\caption{Augmentation of the six-node network. Each box groups six
node equations. Every added node has its own input and measurement;
the horizontal arrows specify the connections between the groups.}
\label{fig:augmentation}
\end{figure}
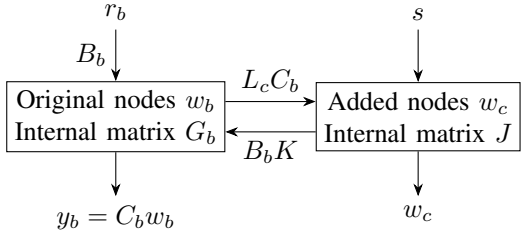

Let \(w_b\in\mathbb C^6\) collect the original node amplitudes.
Add six nodes with amplitudes \(w_c\in\mathbb C^6\), all excited and measured.
Their purpose is to change the complete network spectrum while
preserving the two base solutions.\aicite
Let \(r_b\in\mathbb C^3\) and \(s\in\mathbb C^6\) be the inputs
at the original and added nodes, and \(y_b=C_bw_b\) the original
measurements.
Let \(K\in\mathbb C^{3\times6}\), \(L_c\in\mathbb C^{6\times3}\), and
\(J\in\mathbb C^{6\times6}\) specify the additional connections.
Define the augmented network and its selections by
\begin{equation}\label{eq:augmentation}
\begin{split}
 \mathcal G&=\begin{pmatrix}G_b&B_bK\\L_cC_b&J\end{pmatrix},\\
 \mathcal B&=\begin{pmatrix}B_b&0\\0&I_6\end{pmatrix},\qquad
 \mathcal C=\begin{pmatrix}C_b&0\\0&I_6\end{pmatrix}.
\end{split}
\end{equation}
This again has the form \eqref{eq:network}; its twelve node amplitudes
are the six original amplitudes \(w_b\) followed by \(w_c\).
The matrices \(K,L_c,J\) connect the added nodes to the original
excited and measured nodes and to each other.
Denote its transfer matrix by
\begin{equation}\label{eq:H}
 H=\mathcal C(I_{12}-\mathcal G)^{-1}\mathcal B.
\end{equation}
Figure~\ref{fig:augmentation} groups the nodes and added connections.

\begin{lemma}\label{lem:augmentation}
Suppose \(I_6-G_b\), \(X\), and \(I_{12}-\mathcal G\) are invertible.
Then \(H\) is invertible and
\begin{equation}\label{eq:Hinv}
 H^{-1}=\begin{pmatrix}F^{-1}&-K\\-L_c&I_6-J\end{pmatrix}.
\end{equation}
Consequently \(H\) determines each entry of \(K,L_c,J\), and, on this
domain, equality of augmented transfer matrices is equivalent to equality
of \(F,K,L_c,J\).
Every well-posed candidate with this \(H\) also has invertible
\(X\) and \(I_6-G_b\). If the original base parameters belong to
\(\mathcal D_b\), the augmented fiber consists of exactly the two
base solutions with the same \(K,L_c,J\).
\end{lemma}

\begin{proof}
The node equations give
\[
 y_b=F(r_b+Kw_c),\qquad w_c=L_cy_b+Jw_c+s.
\]
Thus
\[
 \begin{pmatrix}r_b\\s\end{pmatrix}
 =\begin{pmatrix}F^{-1}&-K\\-L_c&I_6-J\end{pmatrix}
  \begin{pmatrix}y_b\\w_c\end{pmatrix}.
\]
By well-posedness, \((y_b,w_c)\) is obtained from \((r_b,s)\) through
the square matrix \(H\). The displayed matrix is therefore a left
inverse of \(H\), proving \eqref{eq:Hinv}.

For any well-posed candidate producing this \(H\), suppose
\(X\xi=0\) for some nonzero \(\xi\in\mathbb C^3\).
The node amplitudes \((\xi,0,0)\), in groups of sizes three, three,
and six, give zero output for input \(((I_3-A)\xi,0)\).
Well-posedness makes this input nonzero, contradicting invertibility
of \(H\). Thus \(X\) is invertible. Eliminating the first three
node amplitudes makes the upper left block of \(H^{-1}\) equal to
\((I_3-A)X^{-1}(I_3-D)\). That block is the original invertible
\(F^{-1}\), so \(I_3-A\) and \(I_3-D\) are also invertible.
Every candidate therefore satisfies the lemma's assumptions.

Proposition~\ref{prop:base} leaves exactly two base candidates on
\(\mathcal D_b\). For either, \eqref{eq:Hinv} determines
\((y_b,w_c)\) from the inputs, and \(X^{-1}(I_3-D)y_b\) gives
the first three node amplitudes. Both augmented candidates are
therefore well posed.
\end{proof}

This identity also preserves the rank information needed for local
identifiability. If \(q_c\) entries of \(K,L_c,J\) are independently
unknown and the remaining entries are fixed zeros, their derivatives
occupy positions in \eqref{eq:Hinv} disjoint from those of \(F^{-1}\).
The augmented Jacobian rank is therefore \(q_c+9\) wherever the base
Jacobian has rank nine. The choice below makes both compatible
augmented matrices satisfy the spectral-radius restriction.
Moreover, \eqref{eq:identities} shows that \(T_b\) changes every
base coefficient on \(\mathcal D_b\). Knowing any one of these
coefficients selects one of the two solutions, for both constructions.

\section{Ambiguity With Arbitrarily Small Spectral Radius}
\label{sec:main}

Use superscript \(T\) for transpose. Define the cyclic permutation
matrix \(U\), and then \(V\), by
\[
 U=\begin{pmatrix}0&0&1\\1&0&0\\0&1&0\end{pmatrix},
 \qquad V=U+U^T.
\]
These matrices satisfy
\(U^3=I_3\) and \(U^T=U^2\).\aicite
For real \(t>0\) and \(x\ne0\), take
\(A=-tU,\ D=-tU^T,\ X=xI_3\). Write \(F(t,x)\) for the
corresponding transfer matrix \eqref{eq:F}. Then
\begin{equation}\label{eq:symmetric}
 F(t,x)^{-1}=x^{-1}\big[(1+t^2)I_3+tV\big],
\end{equation}
so \(F(t,1)=F(t^{-1},t^{-2})\). The two base matrices coincide
at \(t=1\); denote this common matrix by \(G_{b0}\).

We design the added connections at this common point. Seek
\(K_0,L_0\) such that
\(E_+=G_{b0}+B_bK_0\) and \(E_-=G_{b0}-L_0C_b\) have zero
squares, and set
\[
 J_0=G_{b0}+B_bK_0-L_0C_b.
\]
This choice makes the equations in coordinates \((w_b,w_b-w_c)\)
block upper triangular with diagonal blocks \(E_+,E_-\).
Their zero squares make the complete matrix's fourth power zero.
Using \(U^3=I_3\), suitable choices are
\begin{equation}\label{eq:controller}
 K_0=\begin{pmatrix}V&-U\end{pmatrix},\quad
 L_0=\begin{pmatrix}U^T\\-V\end{pmatrix},\quad
 J_0=\begin{pmatrix}U^T&-V\\I_3&U\end{pmatrix}.
\end{equation}
Combining them with the varying base gives
\begin{equation}\label{eq:M}
 M(t,x)=
 \begin{pmatrix}
 -tU&0&V&-U\\
 xI_3&-tU^T&0&0\\
 0&U^T&U^T&-V\\
 0&-V&I_3&U
 \end{pmatrix}.
\end{equation}
All blocks in this display have size \(3\times3\).
Use the nonzero positions in \(M(t,x)\) as the permitted edge set
for \(\mathcal G\). This pattern does not depend on the chosen
nonzero values of \(t,x\).
Assign a separate independent unknown coefficient to every permitted
position. There are nine in \(G_b\), nine in \(K\), nine in \(L_c\),
and fifteen in \(J\), giving \(m=42\).
Thus the repeated numerical values in \eqref{eq:M} specify particular
parameter choices in a class with forty-two independent complex coefficients.

The selections \(\mathcal B,\mathcal C\) in \eqref{eq:augmentation}
excite nodes 1--3 and 7--12 and measure nodes 4--12.
There are no diagonal edges.
Let \(\theta\in\mathbb C^{42}\) collect the independent coefficients
of this structure, and apply \eqref{eq:domain} to \(\mathcal G(\theta)\).

\begin{theorem}\label{thm:main}
For every \(0<\varepsilon\le1\), this twelve-node structure is
generically locally identifiable in \(\Omega_\varepsilon\).
There is a nonempty open set \(W_\varepsilon\subset\Omega_\varepsilon\)
such that every \(\theta\in W_\varepsilon\) has a distinct alternative
\(\widetilde\theta\in\Omega_\varepsilon\) with the same transfer matrix \(H\).
Both alternatives are locally identifiable. Consequently, the structure
is not generically globally identifiable in \(\Omega_\varepsilon\).
The same conclusions hold on the real parameter domain
\(\Omega_\varepsilon\cap\mathbb R^{42}\).
\end{theorem}

\begin{proof}
We first give equal transfer matrices, then make both spectral radii small,
and finally establish ambiguity on an open set of all forty-two
parameters.

\emph{Equal transfer matrices:}
The two augmented matrices
\begin{equation}\label{eq:pair}
 \mathcal G_1(t)=M(t,1),\qquad
 \mathcal G_2(t)=M(t^{-1},t^{-2})
\end{equation}
use the same \(K_0,L_0,J_0\). By Lemma~\ref{lem:augmentation}, their
transfer matrices agree whenever both augmented networks are well posed.
They are distinct for \(t\ne1\).

\emph{Small spectral radii:}
At \(t=1\), both matrices in \eqref{eq:pair} equal \(M(1,1)\).
The design above gives
\begin{equation}\label{eq:nilblocks}
\begin{split}
 E_+&=G_{b0}+B_bK_0
       =\begin{pmatrix}U^T&-U\\I_3&-U^T\end{pmatrix},\\
 E_-&=G_{b0}-L_0C_b
       =\begin{pmatrix}-U&-U^T\\I_3&U\end{pmatrix}.
\end{split}
\end{equation}
Changing coordinates from \((w_b,w_c)\) to \((w_b,w_b-w_c)\) makes
\(M(1,1)\) similar to the block upper triangular matrix
\begin{equation}\label{eq:triangular}
 \begin{pmatrix}E_+&-B_bK_0\\0&E_-\end{pmatrix}.
\end{equation}
The identities for \(U\) imply \(E_+^2=E_-^2=0\).
Squaring \eqref{eq:triangular} therefore gives a matrix whose only
possibly nonzero block is the upper right block; its square is zero.
Consequently
\begin{equation}\label{eq:nilpotent}
 M(1,1)^4=0.
\end{equation}

For a \(d\times d\) matrix \(Q\), let
\(\|Q\|_\infty=\max_{1\le i\le d}\sum_{j=1}^d|Q_{ij}|\), its maximum
absolute row sum.
The standard eigenvalue bound gives
\begin{equation}\label{eq:normbound}
 \rho(Q)^4=\rho(Q^4)\le\|Q^4\|_\infty.
\end{equation}
Both matrices in \eqref{eq:pair} converge to \(M(1,1)\) as \(t\to1\).
Their fourth-power norms consequently tend to zero.
For any prescribed \(0<\varepsilon\le1\), choose \(t\ne1\)
sufficiently close to one that both norms are below \(\varepsilon^4\).
Equation~\eqref{eq:normbound} then places both distinct matrices in
\(\Omega_\varepsilon\), so their transfer matrices are well defined and equal.

\emph{Local identifiability:}
The thirty-three independent added coefficients are determined directly
in \eqref{eq:Hinv}. The base Jacobian has rank nine by
\eqref{eq:rankbase}, since \(\alpha\delta=t^6\) at the first point of
\eqref{eq:pair} and \(\alpha\delta=t^{-6}\) at the second.
Both augmented Jacobians therefore have rank \(33+9=42\).
All their entries are rational functions of the forty-two parameters.
A minor, meaning the determinant of a square submatrix, that is
nonzero at one of these points has a numerator that is a
nonzero polynomial. Outside its zero set, the Jacobian of \(H\)
has full column rank. This establishes generic local identifiability
in \(\Omega_\varepsilon\).

\emph{An open set of ambiguous parameters:}
Extend \(T_b\) to an involution \(\mathcal T\) on the forty-two
parameters by leaving the
thirty-three added coefficients unchanged.
At the first point of \eqref{eq:pair}, one has \(k_i=t^2\),
\(h_i=1+t^2+t^4\), and \(\eta_i=t^2\); thus \(\mathcal T\) maps it to
the second point. All denominators are nonzero there.
The transformation is rational and continuous in a neighborhood, and
preserves \(H\) by Lemma~\ref{lem:augmentation}.

At these two points the fourth-power bounds are strict, the alternatives
are distinct, and both Jacobians have full column rank. These
properties persist in a sufficiently small open neighborhood
\(W_\varepsilon\) of the first point and its image under \(\mathcal T\).
Shrink the neighborhood further, if necessary, so that the base
inverses in the lemma exist and all permitted coefficients remain
nonzero. Then every
\(\theta\in W_\varepsilon\) and its alternative
\(\mathcal T(\theta)\) belong to \(\Omega_\varepsilon\), have equal
transfer matrices, and are distinct and locally identifiable.
The neighborhood is open in \(\mathbb C^{42}\), so it has positive
volume in the real and imaginary coordinates of \(\mathbb R^{84}\).
This disproves generic global identifiability.
The pair is real and \(\mathcal T\) preserves real coefficients.
The neighborhood contains a real open set, and the same nonzero
Jacobian minor establishes generic local identifiability there.
\end{proof}

\subsection{An explicit pair with rational constant coefficients}

For \(\varepsilon=1\), choose \(t=99/100\) in \eqref{eq:pair}.
The common inverse transfer matrix, obtained from \eqref{eq:Hinv}, is
\begin{equation}\label{eq:commoninverse}
 H^{-1}=
 \begin{pmatrix}
 (1+t^2)I_3+tV&-V&U\\
 -U^T&I_3-U^T&V\\
 V&-I_3&I_3-U
 \end{pmatrix}.
\end{equation}
Exact multiplication of the rational matrices gives
\begin{equation}\label{eq:certificates}
\begin{split}
 \|\mathcal G_1(99/100)^4\|_\infty
     &=\frac{33910001}{100000000}<1,\\
 \|\mathcal G_2(99/100)^4\|_\infty
     &=\frac{16976126}{32019867}<1.
\end{split}
\end{equation}
These inequalities certify admissibility through \eqref{eq:normbound}.
Their spectral radii are approximately \(0.509563\) and \(0.404210\),
respectively. The exact equalities \eqref{eq:commoninverse} and
\eqref{eq:certificates} make the pair directly checkable with rational
arithmetic. Theorem~\ref{thm:main} establishes the open-set conclusion
independently of these numerical values.

\subsection{Admissible solution counts and conditioning}

There are also open subsets of \(\Omega_\varepsilon\) on which
the network is globally identifiable relative to that domain.
To see this, choose \(0<t<\varepsilon\) with \(x=1\), initially
setting \(K=L_c=J=0\). The full network and its alternative have
spectral radii \(t\) and \(t^{-1}\), respectively.
Both are well posed and the base parameters lie in \(\mathcal D_b\).
Lemma~\ref{lem:augmentation} gives exactly these two candidates.
Small independent perturbations, with all permitted added entries
nonzero, keep the first radius below \(\varepsilon\) and the second
above it. Thus one admissible solution persists on an open set,
alongside the two-solution regions of Theorem~\ref{thm:main}.
Failure of generic global identifiability therefore does not imply
that almost every admissible network has an admissible alternative.

The spectral-radius bound concerns eigenvalues. At the limit,
\(\rho(M(1,1))=0\) while \(\|M(1,1)\|_\infty=4\).
Along \(M(t,1)\), the reduced Jacobian in \eqref{eq:rankbase} is
\(J_z=I_3-t^2U\). Writing \(\sigma_{\min}\) for the smallest
singular value, the cyclic permutation gives
\[
 \sigma_{\min}(J_z)=|1-t^2|\longrightarrow0\quad(t\to1).
\]
Its inverse therefore amplifies some perturbations by
\(|1-t^2|^{-1}\). The theorem supplies no uniform sensitivity
bound or neighborhood size as \(\varepsilon\) decreases.

\section{Transfer Functions and Conclusions}
\label{sec:conclusion}

\subsection{Algebraic identifiability over rational functions}

Let \(\lambda\) be a symbolic complex frequency variable and let
\(\mathbb K=\mathbb C(\lambda)\) be the field of ratios of polynomials
in \(\lambda\) with complex coefficients. We now replace the scalar
edge coefficients by elements of \(\mathbb K\), keeping the same
network structures and formulas for the transfer matrices. Each coordinate of
\(\theta\in\mathbb K^m\) is then a rational function of \(\lambda\).
Equality of transfer matrices means an identity of rational functions.\aicite

We use the algebraic framework of \cite[Sec.~V]{HMA26}.
For the algebraic calculation, the edge coordinates are independent
variables over \(\mathbb K\). Solutions are counted in an algebraic
closure \(\overline{\mathbb K}\), a field containing \(\mathbb K\)
in which every nonconstant polynomial has a root. An algebraic set is
a set defined by polynomial equations in these coordinates.
An algebraically generic property holds outside a proper algebraic
subset, meaning one smaller than the full parameter domain or the
image of the map \(\theta\mapsto\Phi(\theta)\), as appropriate.
Generic local identifiability in the algebraic sense means that the
generic fiber has dimension zero, equivalently that it consists of
finitely many isolated points. Generic global identifiability requires
one point. Admissibility here excludes zeros of the determinants and
denominators required by the network equations.

\balance
In the notation of \cite{HMA26}, the matrices \(P,Q_r,R_y,S_{yr}\)
define the closed-loop transfer function \(R_yP^{-1}Q_r+S_{yr}\)
from external inputs to measured outputs.
Our augmented network is its specialization
\[
 P=I_{12}-\mathcal G,\qquad Q_r=\mathcal B,\qquad
 R_y=\mathcal C,\qquad S_{yr}=0.
\]

\begin{proposition}\label{prop:field}
Over \(\mathbb K\), the six-node structure \eqref{eq:base} and the
twelve-node structure of Theorem~\ref{thm:main} each have exactly
two points in the generic algebraic fiber. Both are therefore
generically locally identifiable in the algebraic sense and fail
generic global identifiability. If one generic parameter vector has
coordinates in \(\mathbb K\), its distinct alternative also has
coordinates in \(\mathbb K\).
\end{proposition}

\begin{proof}
The identities and solution counts in Proposition~\ref{prop:base} and
Lemma~\ref{lem:augmentation} use only field operations and apply
over \(\overline{\mathbb K}\). They give exactly two solutions on
the nonempty domain \eqref{eq:base-domain}, with well-posedness also
imposed for the augmented network. Formula~\eqref{eq:alternative}
is rational, so it preserves \(\mathbb K\)-valued coordinates;
the added coefficients remain unchanged.
\end{proof}

\subsection{Proper constant networks}

A rational transfer function is \emph{proper} if its numerator
degree does not exceed its denominator degree, and
\emph{strictly proper} if the inequality is strict.
Each real coefficient in \eqref{eq:pair} defines a constant proper
rational transfer function.
The matrix \((I_{12}-\mathcal G)^{-1}\), which maps additive inputs
at all nodes to the node signals, is also constant. It maps bounded
inputs to bounded signals, so the network is stable in this
input-output sense. Both networks have identical transfer matrices
and satisfy the spectral-radius restriction at every frequency.

Algebraically generic uniqueness also fails after imposing properness,
stability, and the spectral-radius bound.
Indeed, \(W_\varepsilon\cap\mathbb R^{42}\) from
Theorem~\ref{thm:main} contains a nonempty real open set of ambiguous
constant networks. This set is Zariski dense over
\(\overline{\mathbb K}\), meaning that no nonzero polynomial in the
forty-two edge coordinates vanishes on all of it.
To see this, take a real open box contained in the set. Fix all but
one coordinate in their respective intervals. A polynomial vanishing
throughout the box has infinitely many roots in the remaining
coordinate, so each of its coefficients is zero. Repeating this
argument for the other coordinates proves that the polynomial is zero.
Thus no proper algebraic exceptional set can contain all these
admissible ambiguous networks.
Their transfer matrices are also Zariski dense in the image of
the map \(\theta\mapsto H(\theta)\).
Indeed, suppose a polynomial vanishes on all these transfer matrices but
not on the full image. Substituting \(H(\theta)\) and clearing
denominators gives a nonzero polynomial that vanishes on the real
open set, a contradiction.

This density statement uses whole transfer functions as coordinates.
In a family parameterized by numerator and denominator coefficients
at a prescribed positive order, constants may be excluded or form
a lower-dimensional subset. The parameter spaces differ.
Preservation of properness, stability, or a prescribed order by
\(\mathcal T\) for arbitrary nonconstant parameters requires a
separate argument. The strictly proper case also remains open.

\subsection{Reproducibility and conclusions}

\ifdefined\verificationdoi
The companion verification archive \cite{verification} contains
code, dependencies, and exact outputs.
\else
Verification code, dependencies, and exact outputs are available
from the author on request.
\fi
The code checks the symbolic involution, nonsymmetric examples, equality
of the transfer matrices, and the power bounds. The Jacobian checks
use all forty-two independent edge positions. It also verifies an
explicit example with only one spectrally admissible solution.
The constructions show that derivative rank tests establish local
recovery without ensuring global uniqueness for this network class,
even under a common spectral-radius bound.

\section*{Acknowledgment}
OpenAI Codex \cite{codex} assisted in developing the construction and
proofs, writing symbolic verification code, and drafting and editing
the abstract and Sections~\ref{sec:intro}--\ref{sec:conclusion}.
The citations in section footnotes identify text drafted with this
assistance.

\end{document}